\documentclass[onecolumn,showpacs,preprintnumbers,amsmath,amssymb,floatfix]{revtex4}
\usepackage[dvips,final]{graphicx}
\usepackage{dcolumn}% Align table columns on decimal point
\usepackage{bm}% bold math
\usepackage{ulem}

\newcommand{\beq}{\begin{equation}}
\newcommand{\eeq}{\end{equation}}

\begin{document}

\title{Mapping the dominant regions of the phase space \\
associated with $\bm{c \bar c}$ production \\
relevant for the Prompt Atmospheric Neutrino Flux}

\author{Victor P.~Goncalves}
\email{barros@ufpel.edu.br}
\affiliation{Instituto de F\'{\i}sica e Matem\'atica,  Universidade
Federal de Pelotas (UFPel), \\
Caixa Postal 354, CEP 96010-900, Pelotas, RS, Brazil}

\author{Rafa{\l} Maciu{\l}a}
\email{Rafal.Maciula@ifj.edu.pl}
\affiliation{Institute of Nuclear Physics PAN, PL-31-342 Cracow, Poland}

\author{Roman Pasechnik}
\email{Roman.Pasechnik@thep.lu.se}
\affiliation{Department of Astronomy and Theoretical Physics,
Lund University, 22362 Lund, Sweden}

\author{Antoni Szczurek\footnote{Also at University of Rzesz\'ow, PL-35-959 Rzesz\'ow, Poland}}
\email{Antoni.Szczurek@ifj.edu.pl} 
\affiliation{Institute of Nuclear Physics PAN, PL-31-342 Cracow, Poland}

\begin{abstract}
\vspace{0.3cm} 
We present a detailed mapping of the dominant kinematical domains contributing to the prompt 
atmospheric neutrino flux at high neutrino energies by studying its sensitivity to the cuts 
on several kinematical variables crucial for charm production in cosmic ray scattering in the atmosphere. 
This includes the maximal center-of-mass energy for proton-proton scattering, the longitudinal momentum 
fractions of partons in the projectile (cosmic ray) and target (nucleus of the atmosphere), the Feynman $x_F$ 
variable and the transverse momentum of charm quark/antiquark. We find that the production of neutrinos 
with energies larger than $E_{\nu} >$ 10$^7$ GeV is particularly sensitive to the center-of-mass energies
larger than the ones at the LHC and to the longitudinal momentum fractions in the projectile 10$^{-8}$
$< x <$ 10$^{-5}$. Clearly, these are regions where we do not control the parton, in particular gluon, densities.
We also analyse the characteristic theoretical uncertainties in the charm production cross section coming 
from its QCD modelling. The precision data on the prompt atmospheric neutrino flux can efficiently 
constrain the mechanism of heavy quark production and underlying QCD dynamics in kinematical ranges 
beyond the reach of the current collider measurements.
\end{abstract}


\maketitle

%%%%%%%%%%%%%%%%%%%%%%%%%%%%%
\section{Introduction}
\label{Sec:Intro}
%%%%%%%%%%%%%%%%%%%%%%%%%%%%%

The recent detection of ultra-high energy neutrino events with deposited
energies up to a few PeV by the IceCube Observatory sets the begining
of neutrino astronomy \cite{IceCube_Science,Aartsen:2014gkd,Aartsen:2016xlq} 
(for a review of IceCube potential for neutrino astronomy, see 
e.g.~Ref.~\cite{Halzen:2010yj}). It is mandatory to know the flux
of atmospheric neutrino produced in cosmic-ray interactions with nuclei
in Earth's atmosphere at different energies with high precision as 
an unavoidable background for cosmic neutrino studies. 
In recent years, the atmospheric high-energy neutrino flux 
became accessible to the experimental studies and, in particular, 
was constrained by several neutrino observatories 
\cite{Abbasi:2010ie,Aartsen:2012uu,Adamson:2012gt,Fukuda:1998ub}.

The available data indicate that the neutrino flux observed in the experiment 
is dominated at low energies ($E_{\nu}\lesssim 10^5$ GeV)  by atmospheric 
neutrinos that arise from the decay of light mesons (pions and kaons), 
denoted as the {\it conventional} atmospheric neutrino flux 
\cite{Honda:2006qj,Barr:2004br,Gaisser:2014pda} while the data for the higher 
energies ($E_{\nu}\gtrsim 10^7$ GeV) are most probably associated with 
cosmic neutrinos.  In the intermediate energy range (10$^{5}$ GeV $< E_{\nu} <$ 
10$^{7}$ GeV), it is expected that the {\it prompt} atmospheric neutrino flux 
associated with the decay of heavy flavoured hadrons, composed of heavy quarks,
become important \cite{ingelman,Martin:2003us,sigl}. In particular, 
it is typically considered that this contribution dominates the atmospheric 
neutrino flux for large neutrino energies ($E_{\nu} > 10^{6}$ GeV). 

This expectation can be easily understood. The increasing competition
between the interaction and decay lengths for pions and kaons at high
energies, implies a reduction of the neutrino flux associated with
decays of these particles. This behaviour is related to the fact that 
the long-lived high-energy light mesons interact and lose their energy 
before decaying into neutrinos. In contrast, in the case of heavy hadrons, 
they have short lifetimes and decay into neutrinos almost immediately 
after their production. Consequently, at very high energies, the atmospheric
neutrino flux is expected to arise from semi-leptonic decays of 
heavy, in particular charmed, hadrons.

Thus, the precise knowledge of the prompt atmospheric flux is 
crucial for the determination of the cosmic neutrino flux. 
This subject has been a theme of intense debate in the literature,
mainly due to the fact that the calculation requires good knowledge 
of the heavy quark production cross section at high energies. 
In the last two years, results of many calculations of this flux were 
presented 
\cite{enb_stasto1,sigl,rojo1,enb_stasto2,rojo2,halzen,brodsky,prosa,sigl2},
focusing on the determination of the theoretical uncertainties 
present in the QCD calculations. These uncertainties are typically associated, 
for example, with the choice of the heavy quark masses, factorization and 
renormalization scales, as well as the contribution of higher order corrections, 
the choice of the parton distribution functions (PDFs) and the treatment of 
QCD dynamics at high energies (very small $x$). The overall theoretical
uncertainty in QCD predictions of the prompt neutrino flux has been
estimated to be a factor of three or a bit larger in Ref.~\cite{Martin:2003us}.
The impact of nuclear effects, saturation and low-$x$ resummation 
was studied in detail in Ref.~\cite{enb_stasto2}. The recent LHC
data for the prompt heavy quark production cross sections 
(see e.g. Refs.~\cite{Aaij:2013mga,Aaij:2015bpa}) significantly reduced
some of these uncertainties with direct impact on the predictions 
for the prompt neutrino flux. However, several questions still remain open. 

The prompt neutrino flux is usually calculated using the semi-analytical
$Z$-moment approach, proposed many years ago in Ref.~\cite{ingelman} and
discussed in detail e.g.~in Refs.~\cite{ers,rojo1}. One of the main
inputs in this approach is the Feynman $x_F$ distribution for 
the heavy quark production in hadronic collisions. As discussed 
e.g.~in Refs.~\cite{prs,ers}, it is expected that the main contribution 
to the prompt neutrino flux comes from large values of $x_F$, that 
are associated with the heavy quark production at forward rapidities. 
Moreover, the production of neutrinos at a given neutrino energy 
$E_{\nu}$ is determined by collisions of cosmic rays with nuclei in
the atmosphere at energies that are a factor of order 100-1000 larger.
One also has that the prompt neutrino flux measured in the kinematical 
range that is probed by the IceCube Observatory and future neutrino telescopes 
is directly associated with the treatment of the heavy quark cross section 
at high energies. Currently, different experiments at the LHC probe 
a limited range in rapidity. In particular, they do not cover rapidities 
larger than 4.5, which corresponds to relatively small values of $x_F \lesssim 0.1$. 
Therefore, the $D$-meson production in the kinematical range of large $x_F$
values is not covered by the LHC detectors.

The main motivation of the current study is to clarify the kinematical range of energies 
and rapidities in the heavy quark production that determine the prompt atmospheric 
neutrino flux in the range probed by the IceCube Observatory. Such an aspect is 
fundamental if we would like to reduce the current theoretical uncertainties. 
Moreover, as the IceCube 2 program \cite{IceCube2} is expected to measure 
neutrinos with energies that are three orders of magnitudes larger than 
the current coverage, it also will help to define what are the theoretical 
issues that should be resolved in order to obtain realistic predictions 
for the future neutrino telescopes. 

In this paper, we concentrate on $c \bar c$ production to understand to
which extend the calculated prompt neutrino flux is reliable.
We therefore neglect the $b \bar b$ production as well as nuclear
effects. The $b \bar b$ component gives about 10\% contribution
to the corresponding $x_F$-distribution and, thus, to the neutrino 
flux \cite{enb_stasto2}.

The paper is organized as follows. In the next section we present a brief review 
of the $Z$-moment formalism for the calculation of the prompt atmospheric 
neutrino flux. Moreover, we describe the main assumptions of our analysis 
and present a comparison of our results and those obtained by the Prosa 
Collaboration \cite{prosa}. In Section~\ref{results} we discuss the different 
cuts assumed in the calculations and analyse their impact on the neutrino flux, 
focussing on $E_{\nu} >$ 10$^6$ GeV. Finally, in Section~\ref{conc} we summarize 
our main conclusions.

%%%%%%%%%%%%%%%%%%%%%%%%%%%%%%%%%%%%%%%%%%%%%%%%%%%%%%%%%%%
\section{Prompt Atmospheric Neutrino Flux}
\label{sec:prompt}
%%%%%%%%%%%%%%%%%%%%%%%%%%%%%%%%%%%%%%%%%%%%%%%%%%%%%%%%%%%

In order to determine the prompt atmospheric neutrino flux at the
detector level we should describe the production and decay of the
heavy hadrons as well as the propagation of the associated particles
through the atmosphere. The evolution of the inclusive particle fluxes
in the Earth's atmosphere can be obtained using the $Z$-moment
approach \cite{ingelman}. In this approach, a set of coupled cascade 
equations for the nucleons, heavy mesons and leptons (and their antiparticles) 
fluxes is solved, with the equations being expressed in terms of the nucleon-to-hadron
($Z_{NH}$), nucleon-to-nucleon ($Z_{NN}$), hadron-to-hadron  ($Z_{HH}$) 
and hadron-to-neutrino ($Z_{H\nu}$) $Z$-moments. For a detailed
discussion of the cascade equations, see e.g.~Refs. \cite{ingelman,rojo1}. 
These moments are inputs in the calculation of the prompt neutrino flux
associated with production of a heavy hadron $H$ and its decay into a
neutrino $\nu$ in the low- and high-energy regimes, which are given, 
respectively, by \cite{ingelman}
\begin{eqnarray}
\phi_{\nu}^{H,low} & = & \frac{Z_{NH}(E) \, Z_{H\nu}(E)}{1 - Z_{NN}(E)} \phi_N (E,0) \,, \\
\phi_{\nu}^{H,high} & = & \frac{Z_{NH}(E) \, Z_{H\nu}(E)}{1 - Z_{NN}(E)}\frac{\ln(\Lambda_H/\Lambda_N)}
{1 - \Lambda_N/\Lambda_H} \frac{m_H c h_0}{E \tau_H} f(\theta) \, \phi_N (E,0) \,,
\end{eqnarray}
where  $H = D^0, D^+, D_s^+$, $\Lambda_c$ for charmed hadrons, $\phi_N(E,0)$ is a primary 
flux of nucleons in the atmosphere, $m_H$ is the decaying particle's mass, $\tau_H$ is the proper 
lifetime of the hadron, $h_0 = 6.4$ km, 
$f(\theta) \approx 1/\cos \theta$ for $\theta < 60^o$, and the effective 
interaction lengths $\Lambda_i$ are given by $\Lambda_i = \lambda_i/(1 -
Z_{ii})$, with $\lambda_i$ being the associated interaction length ($i = N,H$). 
The expected prompt neutrino flux in the detector can be estimated
using the geometric interpolation formula
\begin{eqnarray}
\phi_{\nu} = \sum_H \frac{\phi_{\nu}^{H,low} \cdot  \phi_{\nu}^{H,high}} {\phi_{\nu}^{H,low} + \phi_{\nu}^{H,high}} \,\,.
\label{eq:flux}
\end{eqnarray}
In what follows, we will focus on vertical fluxes ($\theta = 0$) and
assume that the cosmic ray flux $\phi_N$ can be described by a broken 
power-law spectrum \cite{prs}, with the incident flux being
represented by protons ($N = p$). Moreover, we will assume that the charmed 
hadron $Z$-moments can be expressed in terms of the charm $Z$-moment 
as follows: $Z_{pH} = f_H \times Z_{pc}$, where $f_H$ is the fraction of charmed
particle which emerges as a hadron $H$. As in Ref.~\cite{ers}, 
we will assume that $f_{D^0} = 0.565$, $f_{D^+} = 0.246$, 
$f_{D_s^+} = 0.080$ and $f_{\Lambda_c} = 0.094$.

It is important to emphasize that the composition of the particle
content of the ultra high energy cosmic rays in the region beyond 
the ankle ($E \approx 5 \times 10^9$ GeV) still is an open question and 
no clear consensus exists. As the computation of the prompt atmospheric
neutrino flux requires a folding of the heavy-quark cross section with
the incoming cosmic flux, both aspects increase the uncertainty in the
predictions for the flux in the high-energy regime. 
This point has been recently discussed in detail in 
Refs.~\cite{enb_stasto2,prosa}. 

The charm $Z$-moment at high energies can be expressed by 
\begin{eqnarray}
Z_{pc} (E) =  \int_0^1 \frac{dx_F}{x_F} \frac{\phi_p(E/x_F)}{\phi_p(E)} 
\frac{1}{\sigma_{pA}(E)} \frac{d\sigma_{pA \rightarrow charm}(E/x_F)}{dx_F} \,\,,
\label{eq:zpc}
\end{eqnarray}
where $E$ is the energy of the produced particle (charm), $x_F$ 
is the Feynman variable, $\sigma_{pA}$ is the inelastic proton-Air
cross section, which we assume to be given as in Ref.~\cite{sigl}, and 
$d\sigma/dx_F$ is the differential cross section for the charm production, 
which we assume to be given by $d\sigma_{pA \rightarrow charm}/dx_F = 2
\, d\sigma_{pA \rightarrow c \bar{c}}/dx_F$. 

We compute the prompt neutrino flux associated with charmed hadrons by
evaluating all quantities entering the different terms of Eq.~(\ref{eq:flux}). 
In our analysis, we closely follow Refs.~\cite{prs,enb_stasto2}.  
In the analysis of Eq.~(\ref{eq:zpc}) we will use the standard QCD
collinear factorization formalism allowing us to calculate the charm 
production cross section \cite{combridge}. In the leading-order 
collinear factorization approach the differential cross section can be 
written as
\begin{eqnarray}
\frac{d \sigma}{d y_1 d y_2 d^2 p_T} =  \frac{1}{16 \pi^{2} \hat{s}} &\times& [ \;
\overline{ | {\cal M}_{g g \to c \bar c} |^2} 
x_1 g(x_1,\mu_f^2) x_2 g(x_2,\mu_f^2) 
+ \sum_f \overline{ | {\cal M}_{q \bar q \to c \bar c} |^2}
         x_1 q_f(x_1,\mu_f^2) x_2 \bar q_f(x_2,\mu_f^2) \nonumber \\
&+& \sum_f \overline{ | {\cal M}_{q \bar q \to c \bar c} |^2}
         x_1 \bar q_f(x_1,\mu_f^2) x_2 q_f(x_2,\mu_f^2)] \; ,
\label{differential_cross_section} 
\end{eqnarray}
where $p_T$ is the heavy quark transverse momentum, and $y_1$ and $y_2$ 
are the charm and anticharm rapidities, respectively.
The distribution in $x_F$ is obtained by an appropriate binning.
The PDFs will be assumed to be given by the CT14 parametrization \cite{ct14} 
and the hard scattering will be estimated at the leading order taking into 
account both $gg \rightarrow c\bar{c}$ and $q \bar{q} \rightarrow c \bar{c}$ 
subprocesses. The contribution of the next-to-leading order corrections for 
the $x_F$-distribution will be taken into account by multiplying our
predictions by an effective $K$-factor that depends on $x_F$, 
as proposed in Ref.~\cite{prs}. We assume $m_c = 1.5$ GeV, the
factorization and renormalization scales are taken as 
$\mu_f^2 = \mu_r^2 = m_T^2 \equiv (p_T^2 + 4 m_c^2)$. 

We will disregard in the present analysis the nuclear effects, in particular, 
shadowing, i.e. we calculate the cross section for collisions on nuclei as
$\sigma_{pA \rightarrow c \bar{c}} = Z \times \sigma_{p p \to c \bar c} + 
N \times \sigma_{p n \to c \bar c} \approx
A \times \sigma_{pp \rightarrow c \bar{c}}$,
where $Z$, $N$ and $A$ are the number of protons, neutrons and nucleons
in the nucleus of the target, respectively.
In practical calculations we take $^{14}N$ nucleus as the most representative
one. A more refine analysis is possible but would shadow our discussion
of the selected issues. 

\begin{figure}[t]
\begin{center}
\includegraphics[scale=0.4]{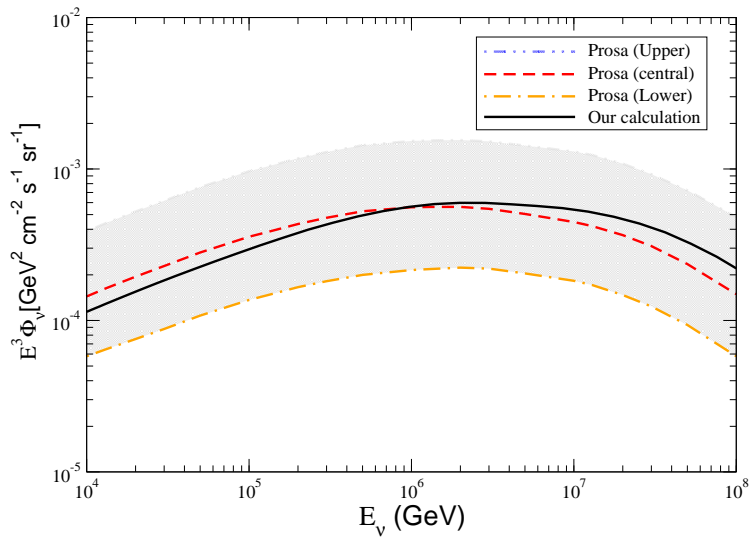}
\caption{Comparison of our predictions for the prompt neutrino flux
  and the Prosa results \cite{prosa}.}
\label{fig:comparison}
\end{center}
\end{figure}

Moreover, we will calculate the effective hadronic interaction lengths
$\Lambda_i$ and the $Z_{pp}$, $Z_{HH}$ and $Z_{H\nu}$-moments as
performed in Ref.~\cite{enb_stasto1}. Although we have done several 
approximations to compute the prompt neutrino flux, our result is similar 
to the central prediction of the Prosa collaboration \cite{prosa}, as shown 
in Fig.~\ref{fig:comparison}. In this figure, we also show the current theoretical 
uncertainty band present in one of the most sophisticated calculations of 
the neutrino flux. Although the available data from collider experiments are 
used in Ref.~\cite{prosa} as an input to constrain the main uncertainties present
in the treatment of heavy quark production, the associated predictions for 
the neutrino flux are still uncertain. The main sources of uncertainty here 
are associated to the modelling of the cosmic ray composition and 
renormalization/factorization scale variations.
%------------------------------------------------------------
%------------------------------------------------------------

%-----------------------
\section{Results}
\label{results}
%-----------------------

In this Section, we wish to understand what is the range of several
kinematical variables relevant for the production of the high energy
neutrinos observed recently by IceCube or for higher energies than 
possible at present. To realize the goal we map the range of several 
kinematical variables such as: center-of-mass energies,
charm transverse momentum ($p_T$), parton momentum fractions 
in the projectile ($x_1$) and target ($x_2$), and the Feynman-$x$ 
($x_F$). All of them determine the size of the cross section and, as 
a consequence, the energy dependence of the prompt neutrino flux.
   
\begin{figure}[t]
\begin{center}
\includegraphics[scale=0.4]{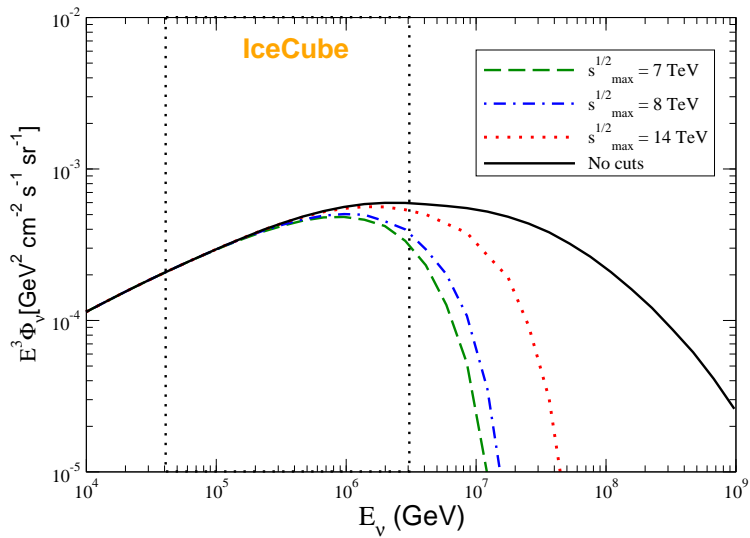}
\caption{Impact of different cuts 
on the maximal  center-of-mass $pp$ collision energy 
for the prompt neutrino flux.}
\label{fig:energy}
\end{center}
\end{figure}

Let us analyze first how the flux of neutrinos from semileptonic decays
of $D$ mesons depends on the maximal center-of-mass collision energy 
included in the calculation. In Fig.~\ref{fig:energy} we present our results 
obtained for different values of the maximal energies considered in 
the analysis of the differential cross section in Eq.~(\ref{eq:zpc}). 
As $x_F$ is integrated and $d\sigma/dx_F$ is probed at the energy
$E/x_F$, one have that $Z_{pc}(E)$ may be influenced by the behaviour 
of distribution at higher energies. In our calculation, we consider three 
different values for the maximum center-of-mass energy allowed in 
the $pp$ collision that generates the heavy quark pair. For comparison 
the full prediction for the flux, denoted as ``no cuts'' in the figure, is presented. 
Here, no energy limitations were imposed.  Moreover, for illustration, the energy 
range probed by the recent IceCube data \cite{Aartsen:2016xlq} is shown as well.
The figure demonstrates that the flux depends on of the cross section for heavy 
quark production in the LHC energy range and at even larger energies.
The latter unexplored region can also have a direct impact on the flux 
at high neutrino energies ($E_{\nu} \ge 10^6$ GeV). 

Moreover, our results indicate that the prompt neutrino flux
for $E_{\nu} \gtrsim 10^7$ GeV is determined by the behaviour of 
the differential cross section in the energy range beyond that 
probed in the Run 2 of the LHC. Consequently, the detection 
of prompt atmospheric neutrinos in this range by the IceCube experiment, 
its upgrade or by other future neutrino telescope, can significantly contribute
to our understanding of several aspects associated with the heavy quark 
production at high energies. Whether we control at present the cross section
for energies above those for the LHC is an open question, 
at least, in our opinion.
%%%%%%%%%%%%%%%%%%%%%%%%%%%
%%%%%%%%%%%%%%%%%%%%%%%%%%%

\begin{figure}[t]
\begin{center}
\begin{tabular}{cc}
\includegraphics[scale=0.43]{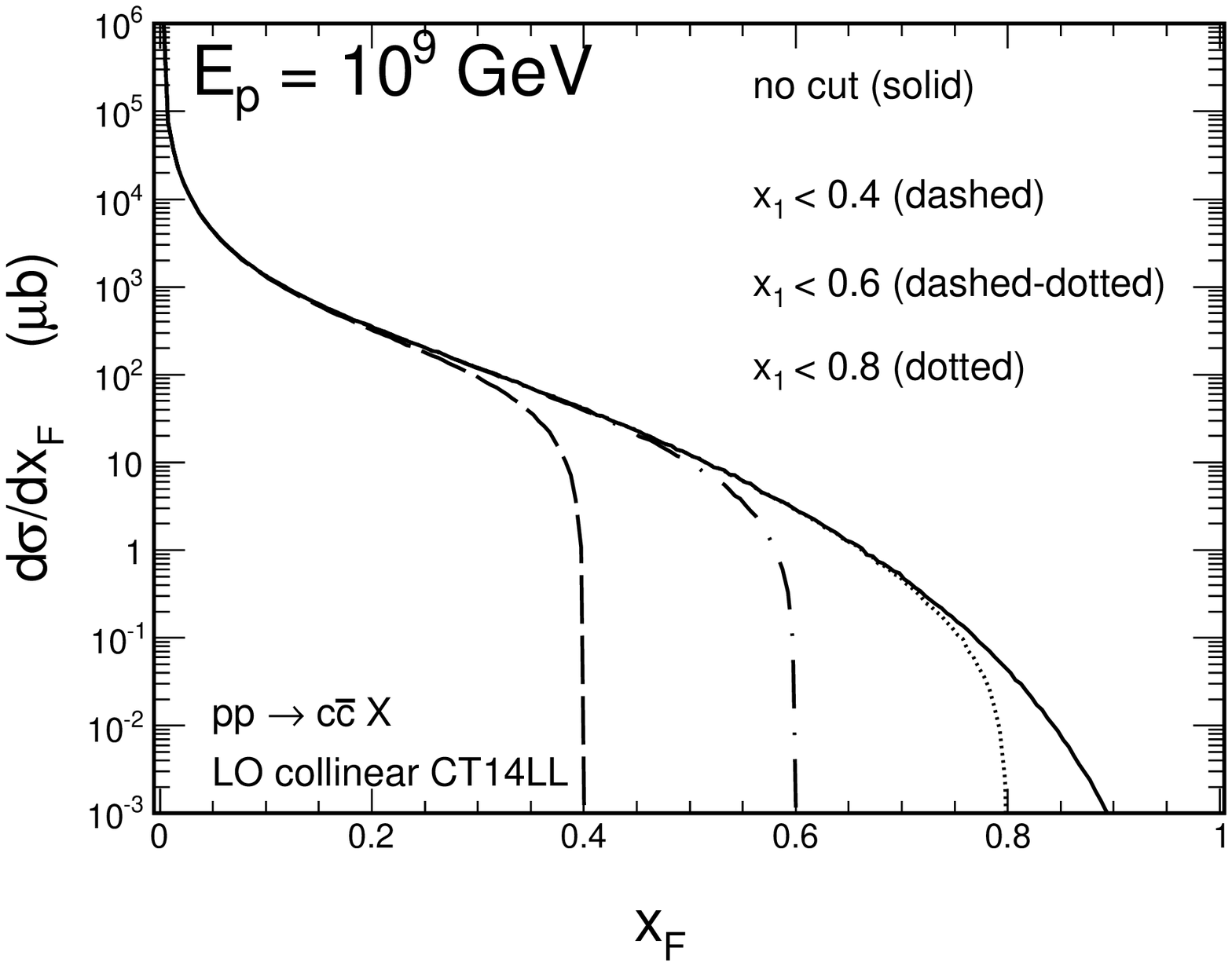}
& \includegraphics[scale=0.35]{xis1cuts2.eps} 
\end{tabular}
\caption{The effect of $x_1$ cuts on the charm production cross 
section $d\sigma/dx_F$ (left) and on the prompt neutrino flux (right).}
\label{fig:xis1}
\end{center}
\end{figure}

In Fig.~\ref{fig:xis1} we present the sensitivity of the charm
production cross section $d\sigma/dx_F$ (left panel) and 
the corresponding energy dependence of the prompt neutrino 
flux on $x_1$ cuts. The notation of the different curves indicate 
the range of $x_1$ values that is included in our calculations. 
The $x_1$ cut has a direct impact on the $x_F$ distribution, 
strongly suppressing the distribution at large $x_F$. Regarding 
the neutrino flux presented on the right panel, we observe that 
the main contribution comes from the intermediate $x_1$-range 
$(0.2 < x_1 <0.6)$. These results demonstrate that the significant 
portion of the neutrino flux comes from very forward (large $x_F$) 
charm production, with the incident parton energy larger than 20\% 
of the projectile nucleon energy at any probed neutrino energy. 

\begin{figure}[t]
\begin{center}
\begin{tabular}{cc}
\includegraphics[scale=0.43]{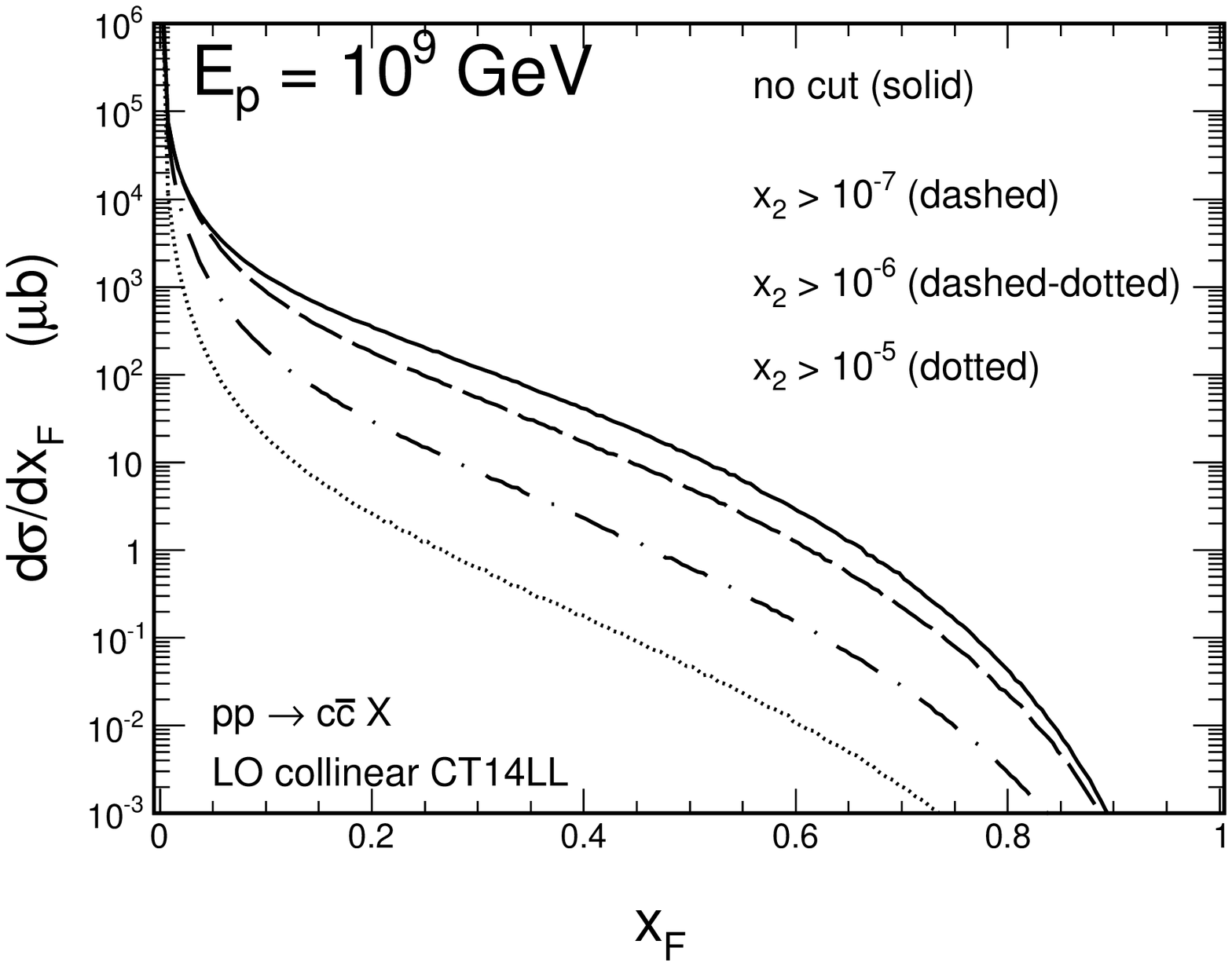}
& \includegraphics[scale=0.35]{xis2cuts2.eps}
\end{tabular}
\caption{The effect of $x_2$ cuts on the charm production 
cross section $d\sigma/dx_F$ (left) and on the prompt neutrino flux (right).}
\label{fig:xis2}
\end{center}
\end{figure}

Analogously, in Fig.~\ref{fig:xis2} we show the corresponding sensitivity 
to the cuts on the target momentum fraction $x_2$. One finds that if the values 
of $x_2 \le 10^{-5}$ are excluded, the $x_F$ distribution gets strongly suppressed 
at intermediate and large $x_F$. In particular, our results indicate that 
the main contribution to the distribution at proton energy $E_p = 10^9$ GeV 
comes from the $10^{-7} < x <  10^{-5}$ range of gluon longitudinal
momentum fractions.

Regarding the neutrino flux, one can see that in the kinematical range probed by 
the recent IceCube data \cite{Aartsen:2016xlq} one observes a strong sensitivity to 
the region of $x_2 <$ 10$^{-5}$. For neutrino energies $E_{\nu} >$ 10$^7$ GeV, 
even the region of $x_2 <$ 10$^{-7}$ becomes important. These values of $x$ 
are beyond those probed by the $pp$ and $ep$ colliders, currently and in the past. 
For instance, the charm production at the LHC (LHCb detector) is sensitive 
to $x_2 >$ 10$^{-5}$, while the HERA data lead to constraints on the
gluon distributions for $x_2 >$ 10$^{-4}$. The smallest values
of $x_2 \sim$ 10$^{-6}$ can be obtained from the inclusive production 
of $\chi_c$ mesons \cite{Cisek:2014ala,Szczurek:2017uvc} in $pp$ collisions 
and in the exclusive $J/\Psi$ photoproduction in hadronic collisions \cite{bruno}. 
However, these possible constraints were not used so far to extract the gluon distributions.
The models of gluon distributions in proton for $x_2 <$ 10$^{-5}$ are therefore rather
uncertain (see e.g.~Ref.~\cite{Ball:2014uwa}). Consequently, the future neutrino telescopes 
will probe the prompt neutrino flux in a weakly explored small-$x$ range of the QCD dynamics.

Very recently, however, in Ref.~\cite{Gauld:2016kpd} the combined set of the LHCb data on $D$-meson 
production at $\sqrt{s}=5,\,7$ and 13 TeV has been shown to constrain the gluon PDF reasonably well 
down to $x\sim 10^{-6}$. Namely, the combined analysis has resulted in an order-of-magnitude
reduction of uncertainties in the gluon PDF compared to such well-known parameterization as 
the NNPDF3.0 \cite{Ball:2014uwa} (for a more recent alternative analysis of the low-$x$ gluon PDF driven 
by the charm LHCb data, see e.g.~Ref.~\cite{deOliveira:2017ega}). Implications of such a reduction
in the gluon PDF uncertainties for the prompt neutrino flux have been discussed in Ref.~\cite{Gauld:2017rbf}.
%%%%%%%%%%%%%%%%%%%%%%%%%%%
%%%%%%%%%%%%%%%%%%%%%%%%%%%
%%%%%%%%%%%%%%%%%%%%%%%%%%%
%%%%%%%%%%%%%%%%%%%%%%%%%%%
%%%%%%%%%%%%%%%%%%%%%%%%%%%
\begin{figure}[t]
\begin{center}
\begin{tabular}{cc}
\includegraphics[scale=0.35]{xfcuts2.eps} & 
\includegraphics[scale=0.37]{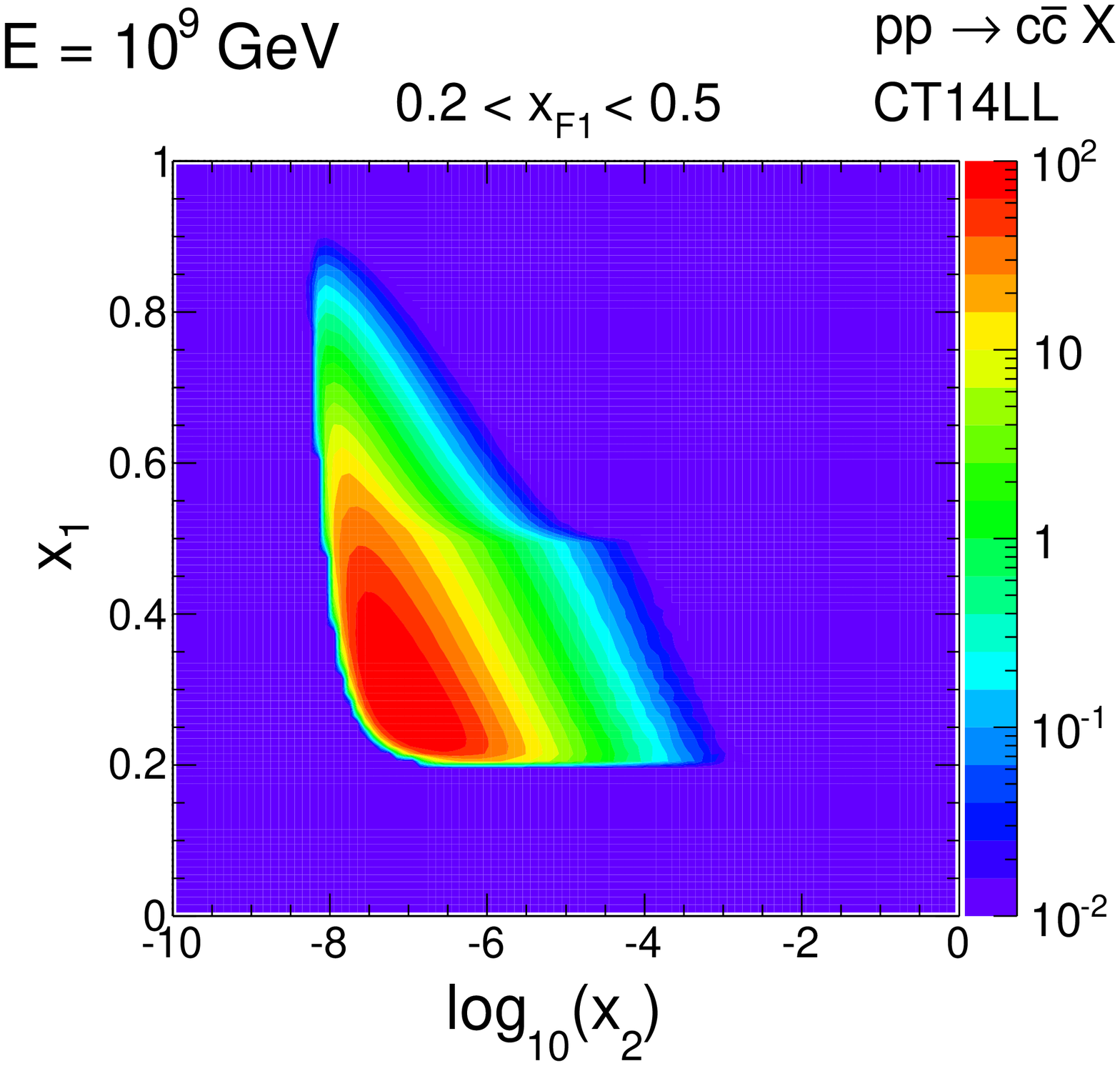}
\end{tabular}
\caption{The effect of cuts on the Feynman variable $x_F$ on the prompt neutrino flux (left), 
and the two-dimensional differential cross section for charm production in $pp$-collisions 
as a function of $x_{1}$ and $\mathrm{log}_{10}(x_{2})$ (right). }
\label{fig:xfpt}
\end{center}
\end{figure}
%%%%%%%%%%%%%%%%%%%%%%%%%%%

In Fig.~\ref{fig:xfpt} (left panel) we present the results for the prompt neutrino flux 
for different cuts on the Feynman $x_F$ variable. We find that the dominant contribution 
to the neutrino flux comes typically from $x_F$ in the region $0.2<x_F<0.5$, which is consistent 
with our previous results for the impact of the $x_1$ and $x_2$ cuts. 

In Fig.~\ref{fig:xfpt} (right panel) we show a two-dimensional plot in $(x_1,\log_{10}x_2)$ for this $x_F$ range. 
For simplicity, in this calculation only the gluon-gluon fusion was
taken into account, which is dominant mechanism 
at large energies (see below). In particular, one can see that the dominant contribution comes from 
the region of $x_1 \in$ (0.2-0.6) and $x_2 \in$ (10$^{-8}$ - 10$^{-5}$). We wish to stress that 
in both these regions of longitudinal momentum fractions gluon distribution is poorly constrained 
(see e.g.~Ref.~\cite{ct14}). The behaviour of the $x_F$ distribution at intermediate $x_F$ is directly 
associated with the charm production at large rapidities, beyond those
probed currently by the LHC detectors. 

%%%%%%%%%%%%%%%%%%%%%%%%%%%
\begin{figure}[t]
\begin{center}
\includegraphics[scale=0.43]{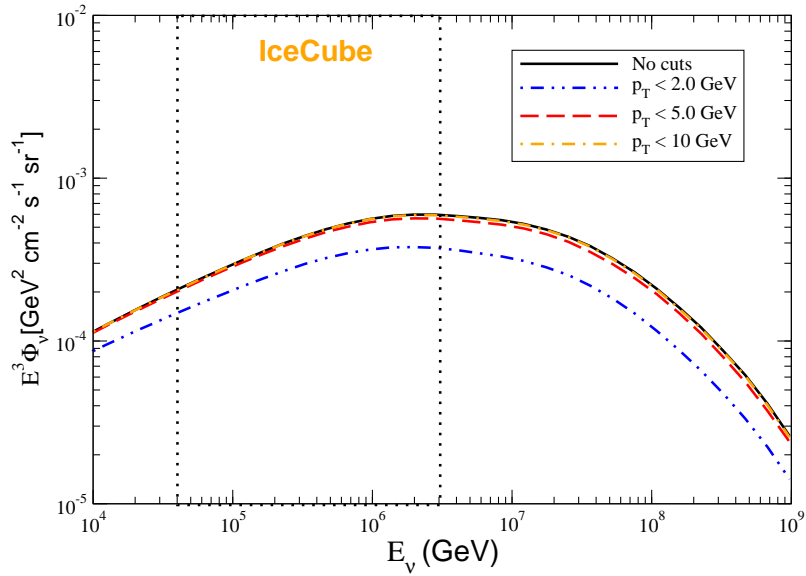}
\caption{The effect of cuts on the quark transverse momentum $p_T$ on the prompt neutrino flux. }
\label{fig:ptcut}
\end{center}
\end{figure}
%%%%%%%%%%%%%%%%%%%%%%%%%%%

For completeness, in Fig.~\ref{fig:ptcut} we analyze the effect of cuts on the quark transverse momentum $p_T$ on the prompt neutrino flux. 
Our results indicate that the prompt neutrino flux is strongly affected by the charm production with tranverse momentum in the $2 < p_T < 5$ GeV range. 
As the description of the transverse momentum spectra for the $D$-meson production at the LHC in this $p_T$ range has a larger theoretical 
uncertainty (see e.g.~Ref.~\cite{prosa}, it also implies a large uncertainty in the neutrino flux predictions in the kinematical range probed by the IceCube.  

\begin{figure}[t]
\begin{center}
\begin{tabular}{cc}
\includegraphics[scale=0.43]{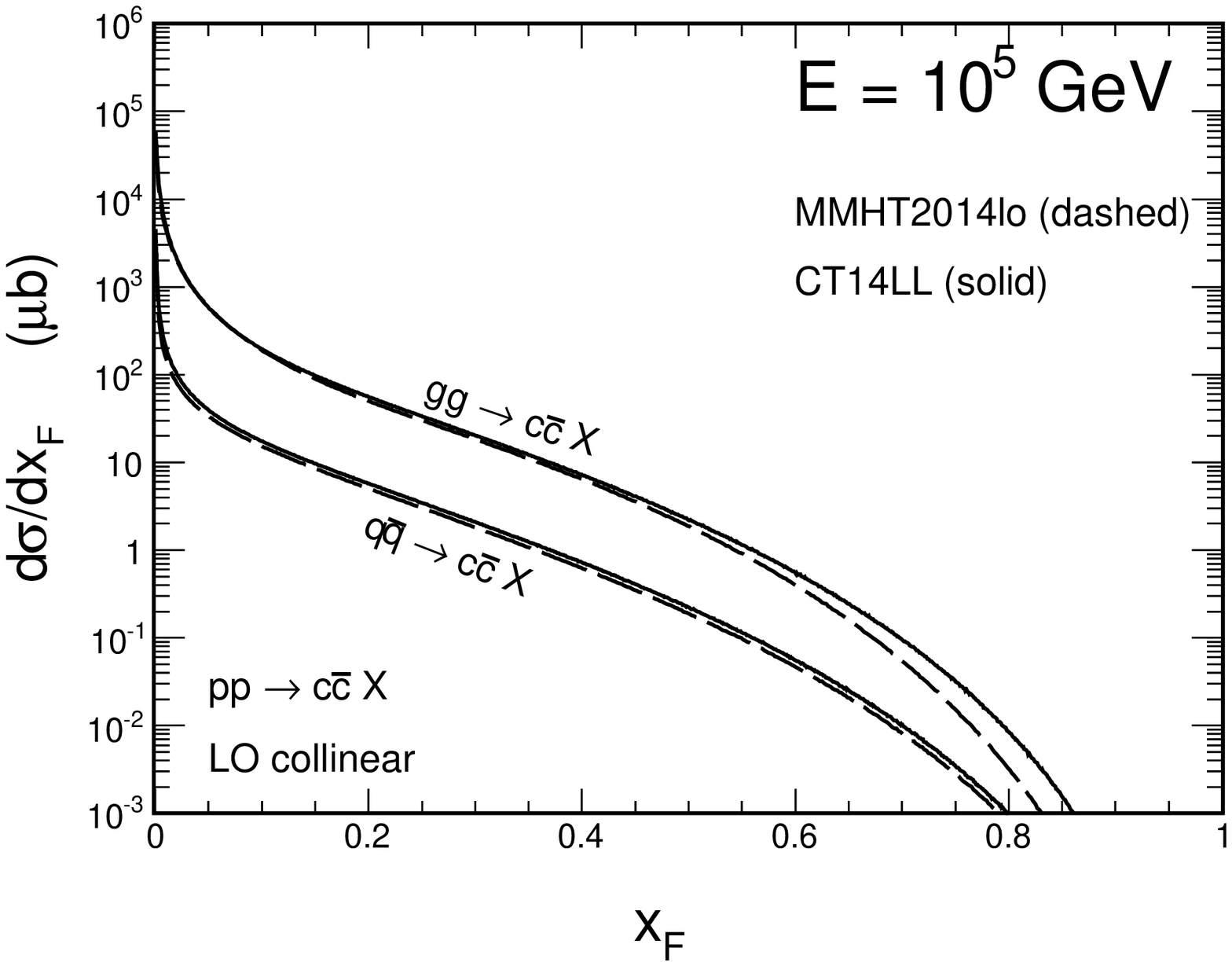} & 
\includegraphics[scale=0.43]{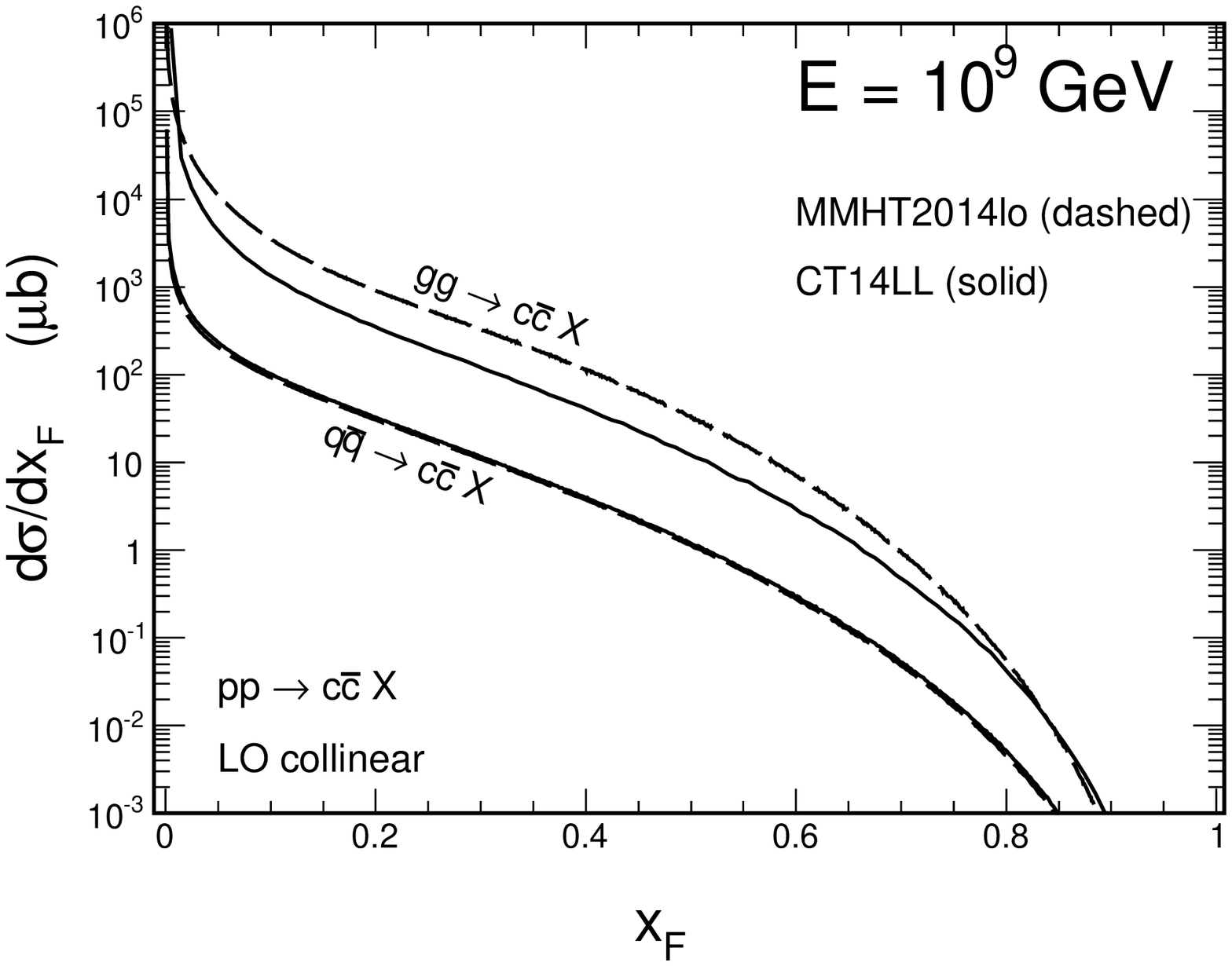}
\end{tabular}
\caption{The charm production cross section $d\sigma/dx_F$ obtained with the leading-order collinear factorization 
for two different energies (left and right panel) and for two different PDF sets. Here, the $gg \to c\bar c$ and 
$q\bar q \to c\bar c$ components are shown separately.}
\label{fig:pdfs_qqbar}
\end{center}
\end{figure}

In order to estimate the sensitive of our predictions on the PDF choice, in Fig.~\ref{fig:pdfs_qqbar} we show the distributions in Feynman $x_F$ for
two different parton distribution sets and for two different energies of the incident cosmic rays, assumed to be protons. In particular, we will compare 
our previous estimates obtained using the CT14LL parametrization \cite{ct14}, with those derived using the MMHT2014LO one \cite{mmht}. 
For completeness, in Fig.~\ref{fig:pdfs_qqbar} we show also the contribution of the quark-antiquark annihilation process. Here, both PDF sets 
give quite similar cross sections for both energies. Regarding to the $gg \rightarrow c \bar{c}$ contribution, one finds that at lower energies (left panel) 
both PDF sets give the same $x_F$-distributions while at higher energies (right panel) they lead to quite different results. Clearly the present experimental data 
obtained at the LHC cannot constrain the gluon distributions at $x <$ 10$^{-5}$. Sinces variations in the $x_F$ distribution at intermediate values of $x_F$ 
have a direct impact on the neutrino flux, we are forced to conclude that the current predictions for the prompt neutrino flux at very high neutrino energies 
are still not reliable.
%%%%%%%%%%%%%%%%%%%%%%%%%%%
%%%%%%%%%%%%%%%%%%%%%%%%%%%

\begin{figure}[t]
\begin{center}
\includegraphics[scale=0.5]{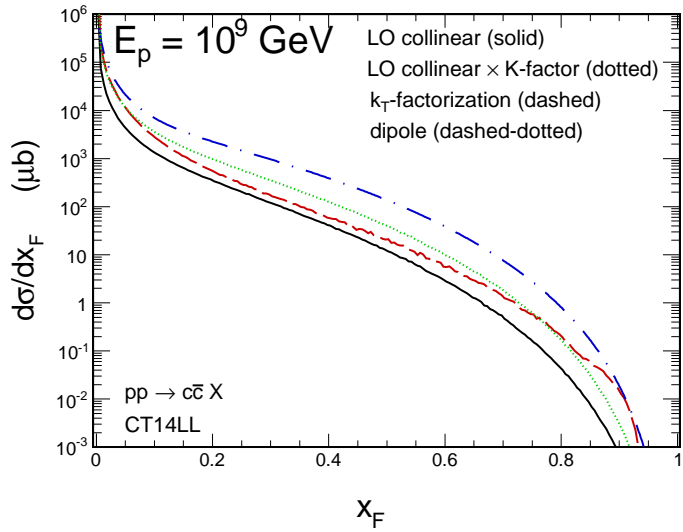}
\caption{The charm production cross section $d\sigma/dx_F$ obtained with three difference QCD approaches: 
 collinear factorization (solid and dotted lines), $k_T$-factorization with the KMR UGDF (dashed line) and the dipole model
presented in Refs.~\cite{vicferHQ,vicroman} (dash-dotted line).}
\label{fig:approaches}
\end{center}
\end{figure}

The description of the QCD dynamics at small-$x$ and the heavy-quark production at large energies and forward rapidities 
are currently the subjects of intense debate. Basically, different formalisms based on different assumptions are able to describe 
the current experimental data. As the behaviour of the prompt neutrino flux at high energies is determined by the $x_F$-distribution 
at intermediate values of $x_F$, it is interesting to compare the predictions of these formalisms for energies probed in neutrino physics. 
In Fig.~\ref{fig:approaches} we compare the charm production cross section obtained in different underlying QCD approaches -- 
the collinear factorization approach (solid and dotted lines), the $k_T$-factorization approach \cite{Gribov:1984tu,Catani:1990xk,Catani:1990eg,
Collins:1991ty} (dashed line) and the dipole model accounting for the saturation phenomena \cite{Kopeliovich:1981pz,nnz,boris} (dash-dotted line). 
These distinct approaches for the heavy quark production in hadronic collisions differ in their basic assumptions and partonic pictures.

While in the collinear framework, all particles involved are assumed to be on mass shell, carrying only longitudinal momenta, and the cross section 
is averaged over two transverse polarizations of the incident gluons, in the $k_T$-factorization approach the Feynman diagrams are calculated 
taking into account the virtualities and all possible polarizations of the incident partons. Moreover, in the $k_T$-factorization approach 
the unintegrated gluon distributions are employed instead of the usual
collinear distributions.

%%%%%%%%%%%%%%%%%%%%%%%%%%%
%%%%%%%%%%%%%%%%%%%%%%%%%%%

In contrast, in the color dipole formalism \cite{Kopeliovich:1981pz,nnz,boris} the basic partonic picture of heavy quark production in gluon-gluon 
interactions is such that, before interacting with the hadron target, a gluon is emitted by a projectile and fluctuates into a color octet pair $Q\bar{Q}$, 
its lowest-order Fock component. The dipole approach does not rely on QCD factorisation \cite{Kopeliovich:2005ym} and is based upon 
the universal ingredients such as the dipole cross section and the light-cone wave function for a given Fock component of the projectile that 
undergoes scattering off the target nucleon. One the main motivations to use this approach is that it allows us to take into account the non-linear 
effects in the QCD dynamics, expected to be important at large energies, the QCD factorisation breaking effects at large Feynman $x_F$, as well as 
the higher-order QCD corrections and coherence phenomena (for more details, see e.g.~Refs.~\cite{rauf,Pasechnik:2015fxa,Basso:2015pba,
Goncalves:2016qku,vicroman} and references therein).

The framework of $k_T$-factorization used here was successfully applied by two of us for single \cite{Maciula:2013wg} and double \cite{Maciula:2013kd} 
open charm meson production at the LHC, as well as for leptons from semileptonic decays of heavy mesons at RHIC \cite{Maciula:2015kea}. As a default choice, 
we use the Kimber-Martin-Ryskin (KMR) \cite{Kimber:2001sc,Watt:2003vf} unintegrated gluon distribution functions (uGDFs) that were shown recently 
to effectively include a part of real higher-order corrections in charm production \cite{Maciula:2016kkx}. In the case of the dipole approach, we will consider 
the predictions obtained recently in Refs.~\cite{vicferHQ,vicroman}
which describe the current LHCb data, at least, in the high-$p_T$
domain. As discussed above, here the Feynman $x_F$ distributions are
very sensitive to the very small transverse momenta.

We observe in Fig.~\ref{fig:approaches} a significant order-of-magnitude difference between the predictions of the dipole and collinear QCD approaches, 
with the $k_T$-factorization result being in between. We wish to point out here that the contributions of the gluon bremsstrahlung off light 
$q\to q+(G\to Q\bar Q)$ and heavy $Q(\bar Q)\to Q(\bar Q)+G$ (anti)quarks are not included in the current dipole-model analysis and that are worth 
further exploration. The large cross section in the dipole model is somewhat unexpected as this approach includes saturation effects that should lead rather 
to a reduction of the cross section compared to the traditional collinear factorization approach. One the other hand, a similar effect has been observed at low heavy 
quark (heavy meson) transverse momenta $p_T<m_Q$ where the dipole results overshoot the LHC data on open heavy flavor production \cite{vicroman}. 
Can this effect be caused by an approximate treatment of the kinematics and the dipole cross section or due to the missing higher-Fock (higher-twist) contributions 
in the current dipole-model analysis? Recently, the Drell-Yan process in the dipole picture and the associated kinematic constraints has been thoroughly discussed 
by some of us in Refs.~\cite{Basso:2015pba,Goncalves:2016qku,Schafer:2016qmk} while the higher-twist corrections remain uncertain. A proper analysis of this 
issue for heavy flavor production in the dipole picture is left for a future work.

\begin{figure}[t]
\begin{center}
\includegraphics[scale=0.43]{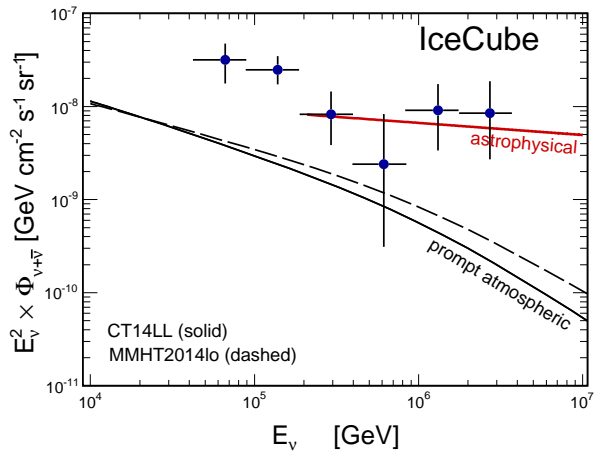}
\caption{Comparison of predictions obtained with the CT14 and MMHT PDFs
  for the prompt neutrino flux.
The data points are taken from Ref.~\cite{Aartsen:2016xlq}. 
For comparison, the fit for the astrophysical contribution, proposed in Ref.~\cite{Aartsen:2016xlq}, 
is presented as well. }
\label{fig:IceCube-data}
\end{center}
\end{figure}

Finally, in Fig.~\ref{fig:IceCube-data} we show the six-year experimental data collected by the IceCube Observatory \cite{Aartsen:2016xlq} 
together with our predictions for the neutrino flux calculated with two different current gluon PDFs. Both theoretical fluxes are below 
the IceCube data but unfortunately we cannot draw at present too strong
conclusions. For comparison, we show a result of a simple fit (unbroken 
power-law isotropic distribution) proposed in
Ref.~\cite{Aartsen:2016xlq}, 
which is consistent with the yet low-statistics data. 

Considering the several aspects discussed above one finds that in order to disentangle the magnitude of the astrophysical contribution to the neutrino flux, 
it is mandatory to get a better theoretical control of the prompt neutrino flux. Although the new experimental data from the LHC will be useful, they will 
not well constrain the charm production and the QCD dynamics in the kinematical ranges that determine the prompt neutrino flux at IceCube and future 
neutrino telescopes. Therefore, the experimental measurement of the neutrino flux and the separation of the prompt contribution are important challenges 
that should be surpassed in order to improve our understanding of strong interactions at high energies as well as of neutrino physics in astrophysical events.  
%%%%%%%%%%%%%%%%%%%%%%%%%%%
%%%%%%%%%%%%%%%%%%%%%%%%%%%

%%%%%%%%%%%%%%%%%%%%%
\section{Summary}
\label{conc}
%%%%%%%%%%%%%%%%%%%%%

One of the current challenges in neutrino physics is to disentangle the signals of astrophysical origin from those associated with 
atmospheric interactions. The precise determination of the conventional and prompt atmospheric neutrino fluxes is fundamental for 
the interpretation of the results from neutrino observatories, such as the IceCube. In the last years, several groups estimated 
the prompt neutrino flux using different theoretical approaches e.g.~for the calculation of the charm production cross section, 
charm fragmentation, cosmic ray flux etc. These studies demonstrated that the theoretical uncertainties are large, although they 
were reduced by the recent collider data and theoretical developments for the heavy quark production. Consequently, it is important 
to map the kinematical range that is probed by high-energy atmospheric neutrinos in order to clearly define the next steps that should be 
performed to obtain precise predictions for the atmospheric neutrino flux. This has been one of the main goals of our current study.

In this paper, we have presented a detailed analysis of the kinematical domains that dominate the charm and prompt atmospheric neutrino 
production in cosmic rays relevant for the IceCube experiment by exploring the sensitivity of the corresponding neutrino flux and the charm 
cross section to the cuts on the maximal $pp$ c.m. energy, the longitudinal momentum fraction in the target and projectile, the Feynman $x_F$ 
and $p_T$ variables included in the calculation. We have found that in order to address production of high-energy neutrinos ($E_{\nu} >$ 10$^7$ GeV) 
one needs to know the charm production cross section for energies larger than those available at the LHC as well as the parton/gluon distributions 
for the longitudinal momentum fractions in the region 10$^{-8}$ $< x <$ 10$^{-5}$. Since this region of $x$ is not available at the collider 
measurements in the moment, the predictions in the collinear factorization approach and the $k_T$-factorization approach are not very reliable. 
If it was possible to disantagle the prompt atmospheric contribution
from the cosmogenic one, it could perhaps become possible to put some contraints 
on the gluon distributions for extremely small longitudinal momentum fractions. This option requires a more dedicated study in the future.

We have also indicated the characteristic theoretical uncertainties in the charm production cross section obtained within different 
QCD approaches typically used by different groups in the analysis of prompt neutrino fluxes such as the leading-order collinear 
factorization approach, $k_T$-factorization and the dipole model accounting for the saturation phenomena.

Our results demonstrate that in order to predict the prompt neutrino flux for typical neutrino energies at the IceCube Observatory 
and future neutrino telescopes, we should extrapolate the behaviour of the heavy quark cross sections and energy distributions 
beyond the range accessible experimentally by current collider measurements. These results indicate that theoretical and experimental 
studies of the prompt atmospheric neutrino flux can provide an important information about the mechanism of heavy quark production 
as well as the description of the QCD dynamics in a kinematical range beyond that reached by the current colliders. At the current stage of 
research, it is premature to decide whether the measurement at the IceCube Observatory can provide a new information on the gluon distribution 
at very low longitudinal fractions $x \sim 10^{-7}$.

\begin{acknowledgments}
V.P.G.~is partially supported by CNPq, Brazil. R.P.~is partially supported by the Swedish Research Council, contract number 621-2013-428 
and by CONICYT grant PIA ACT1406. Rafa{\l} Maciu{\l}a and Antoni Szczurek were partially supported by the Polish National Science Center 
grant DEC-2014/15/B/ST2/02528. A.S. thanks Tomasz Palczewski for expaining some details
about the IceCube Observatory and related physics.
\end{acknowledgments}

%%%%%%%%%%%%%%%%%%%%%

\end{document}